\documentclass[12pt]{iopart}

\usepackage{iopams} 
\usepackage{bm,graphics,graphicx,epsfig,color}

\newcommand{\bv}{{\bm v}}
\newcommand{\bx}{{\bm x}}
\newcommand{\rdot}{\dot{r}}  

\begin{document}

\title[Testing GR with compact-body orbits]{Testing general relativity with compact-body orbits: A modified Einstein-Infeld-Hoffmann framework}

\author{Clifford M. Will$^{1,2}$}

\ead{cmw@phys.ufl.edu}
\address{
$^1$ Department of Physics,
University of Florida, Gainesville FL 32611, USA
 \\
$^2$ GReCO, Institut d'Astrophysique de Paris, UMR 7095-CNRS,
Universit\'e Pierre et Marie Curie, 98$^{bis}$ Bd. Arago, 75014 Paris, France
}

\begin{abstract}
We describe a general framework for analyzing orbits of systems containing compact objects (neutron stars or black holes) in a class of Lagrangian-based alternative theories of gravity that also admit a global preferred reference frame.  The framework is based on a modified Einstein-Infeld-Hoffmann (EIH) formalism developed by Eardley and by Will, generalized to include the possibility of Lorentz-violating, preferred-frame effects.   It uses a post-Newtonian $N$-body Lagrangian with arbitrary parameters that depend on the theory of gravity and on ``sensitivities'' that encode the effects of the bodies' internal structure on their motion.    We determine the modified EIH parameters for the Einstein-{\AE}ther and Khronometric vector-tensor theories of gravity.
We find the effects of motion relative to a preferred universal frame on the orbital parameters of binary systems containing neutron stars, such as a class of ultra-circular pulsar-white dwarf binaries; the amplitudes of the effects depend upon ``strong-field'' preferred-frame parameters $\hat{\alpha}_1$ and $\hat{\alpha}_2$, which we relate to the fundamental modified EIH parameters.    We also determine the amplitude of the ``Nordtvedt effect'' in a triple system containing the pulsar J0337+1715 in terms of the modified EIH parameters.

\end{abstract}


\maketitle

\section{Introduction}
\label{sec:intro}

The $N$-body equations of motion in the post-Newtonian limit of general relativity have been derived by numerous authors \cite{LorentzDroste,1916MNRAS..77..155D,1965npga.book.....L,1964tstg.book.....F}.
A key assumption that
went into those analyses was that the weak-field, slow-motion limit of gravitational theory applied everywhere, in the interiors of the bodies as well
as between them.  This assumption restricted the applicability of the equations of motion to systems such as the solar system.
However, when dealing with systems
containing a neutron star or a black hole with highly relativistic spacetimes near or inside them, one can no longer
apply the assumptions of the post-Newtonian limit everywhere, except
possibly in the interbody region between the relativistic bodies.  Instead,
one must employ a method for deriving equations of motion for compact
objects that, within a chosen theory of gravity, involves solving the full,
relativistic equations for the regions inside and near each body, solving
the post-Newtonian equations in the interbody region, and matching
these solutions in an appropriate way in an ``overlap region'' surrounding
each body.  This matching  leads to constraints on the motions
of the bodies
which constitute the sought-after equations of motion.  In general relativity, a related procedure that avoided explicit inner solutions by exploiting surface integrals around each body  was carried out in the classic work of Einstein, Infeld and Hoffmann
(EIH) \cite{1938AnMat..39...65E}.

In the fully weak-field post-Newtonian (PN) limit, it is also known that the
motion of self-gravitating, post-Newtonian fluid or dust bodies is independent of their internal structure, i.e., there is no Nordtvedt effect \cite{1968PhRv..169.1017N,1971ApJ...163..611W}.  Each body moves on a geodesic of
the post-Newtonian interbody metric generated by the other bodies, with
proper allowance for post-Newtonian terms contributed by its own interbody field.  The equations of motion obtained by this method are identical to the EIH equations.   This structure independence has also been verified explicitly to second post-Newtonian (2PN) order for self-gravitating fluid bodies \cite{2007PhRvD..75l4025M}. It is therefore reasonable to assume that the EIH equations are valid for systems of compact bodies (neutron stars or
black holes).  The only restriction is that they be quasistatic, nearly spherical, and sufficiently small compared to their separations that tidal interactions may be neglected (throughout this discussion, we will ignore the effects of spin).  
This was verified in a restricted context for nonrotating black holes in a seminal paper 
by d'Eath \cite{1975PhRvD..12.2183D}, and was subsequently extended to a variety of contexts  \cite{1978GReGr...9..809R,1978GReGr...9..821R,1980PhRvD..22.1853K,1985PhRvD..31.1815T,1987PhRvD..36.2301A,2000PhRvD..62f4002I,2008PhRvD..78h4016T}.   

Key to this result is the validity of the Strong Equivalence
Principle within general relativity, which
guarantees that the structure of each body is independent of the surrounding gravitational environment.  By contrast, most alternative theories of gravity possess additional gravitational fields, whose values in the
matching region can influence the structure of each body, and thereby can
affect its motion.    Using a cyclic {\em gedanken} experiment within a freely falling frame that encompasses a given body, and that assumes only conservation of energy (see Sec.\ 2.5 of \cite{tegp} for discussion), it can be shown that, if the
binding energy $E_{\rm B}$ of a body varies as a result of a variation in an external parameter or field $\psi$, the body experiences an additional acceleration given by
$
\delta \bm{a}  \sim {m}^{-1} \bm{\nabla} E_{\rm B} (\bm{x}, \bm{v}) \sim (\partial \ln m/\partial \psi) \bm{\nabla} \psi$.
Thus, the bodies need not follow geodesics of any metric, but instead their
motion may depend on their internal structure.

In practice, the EIH-inspired matching procedure  is very cumbersome 
\cite{1975PhRvD..12.2183D}.  Within general relativity, a simpler method for
obtaining the EIH equations of motion is to treat each body as a ``point'' mass of inertial mass $m_a$ and to solve Einstein's equations using a point-mass matter action or energy-momentum tensor, with proper care to neglect or regularize infinite ``self'' fields (at high PN orders, this regularization can become very complicated \cite{2014LRR....17....2B}).   In the action for general relativity, we thus write
\begin{equation}
I = \frac{1}{16 \pi G} \int R \sqrt{-g} \,  d^4x - \sum_a m_a \int d\tau_a \,,
\end{equation}
where $\tau_a$ is proper time along the world line of the $a$th body.  By solving the field equations to 
1PN order, it is then possible to derive straightforwardly from the matter action an $N$-body EIH action in the form
\begin{equation}
I_{\rm EIH} = \int L(\bx_1, \dots \bx_N, \bv_1, \dots \bv_N ) dt \,,
\label{eq10:EIH}
\end{equation}
with a Lagrangian $L$
written purely in terms of the variables $(\bx_a,\, \bv_a)$ of the bodies.  The result is the 1PN $N$-body Lagrangian of general relativity.  The $N$-body EIH equations of motion are then given by 
\begin{equation}
\frac{d}{dt} \frac{\partial L}{\partial v_a^j} -  \frac{\partial L}{\partial x_a^j}  = 0 , \quad a = 1, \dots N \,.
\end{equation}

In alternative theories of gravity, we assume that the only difference is the possible
dependence of the mass on the boundary values of the auxiliary fields.  Thus, following the suggestion of Eardley \cite{1975ApJ...196L..59E}, we merely replace the
constant inertial mass $m_a$ in the matter action with the variable inertial
mass $m_a (\psi_A)$, where $\psi_A$ represents the values of the external auxiliary
fields, evaluated at the  body (we neglect their variation across the interior of the matching region), with infinite self-field contributions
excluded.  
The functional dependence of $m_a$ upon the variable $\psi_A$ will
depend on the nature and structure of the body.  Thus, we write the action of the alternative theory in the form
\begin{equation}
I = I_G - \sum_a \int m_a \left (\psi_A \right ) d\tau_a \,,
\end{equation}
where $I_G$ is the action for the metric and auxiliary fields $\psi_A$.  
In varying the action with respect to the fields $g_{\mu\nu}$ and $\psi_A$, the variation of
$m_a$ must now be taken into account.   In the post-Newtonian limit, where the fields $\psi_A$ are expanded about asymptotic values $\psi_A^{(0)}$ according to
$\psi_A = \psi_A^{(0)} + \delta \psi_A$, it is generally sufficient to expand $m_a (\psi_A)$ in the form
\begin{eqnarray}
m_a (\psi_A) = m_a (\psi_A^{(0)} ) + \sum_A \frac{\partial m_a}{\partial \psi_A^{(0)}} \delta \psi_A + \frac{1}{2} \sum_{A,B} \frac{\partial^2 m_a}{\partial \psi_A^{(0)}\partial \psi_B^{(0)}}
\delta \psi_A\delta \psi_B + \dots \,.
\label{eq10:mexpand}
\end{eqnarray}
Thus, the final form of the metric and of the $N$-body Lagrangian will
depend on $m_a \equiv m_a (\psi_A^{(0)} )$ and on the parameters $\partial m_a /\partial \psi_A^{(0)}$, and so on. 
It is conventional to use
the term ``sensitivities'' to describe these parameters, since they measure the
sensitivity of the inertial mass to changes in the fields $\psi_A$. 
Thus, we 
define
\begin{eqnarray}
s_a^{(A)} &\equiv \frac{\partial \ln m_a}{\partial \ln \psi_A^{(0)}} \,, \quad
{s'_a}^{(AB)} \equiv \frac{\partial^2 \ln m_a}{\partial \ln \psi_A^{(0)} \partial \ln \psi_B^{(0)}}  \,,
\label{eq10:sensitivity0}
\end{eqnarray}
and so on, where the derivatives are typically taken holding the total baryon number of the body fixed, for bodies made of matter.   For black holes, some other method must be used to identify the sensitivities.     Similar sensitivities can be defined for the radius and moment of inertia of the compact body. 
Gralla \cite{2010PhRvD..81h4060G,2013PhRvD..87j4020G} developed a more general theory of the motion of ``small'' bodies characterized by such parameters as mass, spin and charge in an environment of external fields, and
 argued that Eardley's ansatz is a special case of that general framework.

These ideas were incorporated into a ``modified EIH formalism'' in Sec.\ 11.3 of {\em Theory and Experiment in Gravitational Physics} \cite{tegp} (based in part on unpublished notes by Eardley).  However, that formalism was restricted by requiring that the $N$-body Lagrangian be invariant under suitable low-velocity Lorentz transformations.   This precluded the possibility of preferred-frame effects in compact-body dynamics.   A number of developments since that time (1981) have made it desirable to relax that assumption and to create a more general modified EIH formalism.  The first was the formulation of a variety of alternative theories of gravity with auxiliary vector and tensor fields, whose presence establishes a preferred frame of reference where the components of those fields take on privileged values.  Examples include vector-tensor theories such as Einstein-{\AE}ther theory and Khronometric theory \cite{2001PhRvD..64b4028J,2002cls..conf..331M,2010PhRvL.104r1302B,2011JHEP...04..018B}, and scalar-vector-tensor theories, such as TeVeS and STV \cite{2004PhRvD..70h3509B,2008PhRvD..77l3502S,2006JCAP...03..004M}.   A second development was a number of very precise tests of preferred-frame effects in the orbits of ultralow-eccentricity binary pulsar systems \cite{2000ASPC..202..113W,2012CQGra..29u5018S,2013CQGra..30p5020S}.   These analyses quoted bounds on preferred-frame parameters $\hat{\alpha}_1$ and $\hat{\alpha}_2$, where the hats are meant to denote some unspecified generalization of the usual parametrized post-Newtonian (PPN) preferred-frame parameters ${\alpha}_1$ and ${\alpha}_2$ to strong-field situations.   One goal of this paper is to relate these ``hatted'' parameters explicitly to parameters of the modified EIH formalism.  In the special cases of Einstein-{\AE}ther and Khronometric theories, we will obtain the modified EIH parameters explicitly, making it possible to express $\hat{\alpha}_1$ and $\hat{\alpha}_2$ in terms of the fundamental parameters of these theories and of the sensitivities of the compact bodies in the system.  We will also express the ``Nordtvedt'' parameter describing a failure of the universality of free fall in a compact-body system such as the pulsar J0337+1715 in a triple system with two white dwarf companions \cite{2014Natur.505..520R} in terms of parameters of this modified EIH formalism.

In Sec.\ \ref{sec:modifiedEIH}, we describe this generalization of the modified EIH formalism, and in Sec.\ \ref{sec:alternative}, we obtain the parameters of the formalism in a selection of alternative theories.  Section \ref{sec:applications} applies the formalism to two-body and three-body dynamics, obtaining a number of observable effects in terms of the modified EIH parameters.   In Sec.\ \ref{sec:conclusions} we make concluding remarks.   We use units in which the locally-measured gravitational constant $G$ and the speed of light $c$ are unity.  

\section{Modified Einstein-Infeld-Hoffmann formalism}
\label{sec:modifiedEIH}

We construct a general EIH formalism  using arbitrary parameters whose values depend both on the theory under study and on the nature of the bodies in the system.   In this case, however, the parameters appear in the $N$-body 1PN Lagrangian rather than in the metric.
This implies that we are restricting attention to Lagrangian-based metric theories of
gravity, known as semiconservative theories (see Sec.\ 4.4 of \cite{tegp} or \cite{tegp2}  for discussion).
We will also restrict attention to theories that have no
Whitehead term in the post-Newtonian limit; all of the currently popular theories as described in \cite{2014LRR....17....4W} satisfy this constraint.   Our goal is to generalize the formalism presented in \cite{tegp}, which made a restriction to fully conservative theories.  The distinction between the two classes of theories is that semiconservative theories admit preferred-frame effects at 1PN order, i.e. one or more of their PPN parameters $\alpha_1$ and  $\alpha_2$ is nonzero, while fully conservative theories have $\alpha_1=\alpha_2=0$ and do not have preferred-frame effects at 1PN order.    Nordtvedt \cite{1985ApJ...297..390N} developed a similar formalism for compact-body dynamics, but did not address the consequences of preferred-frame effects.

Each body is characterized by an inertial mass $m_a$, defined to be the
quantity that appears in the conservation laws for energy and momentum
that emerge from the EIH Lagrangian.  We then write for the metric,
valid in the interbody region and far from the system,
\begin{eqnarray}
g_{00} &= -1 + 2 \sum_a \alpha^*_a \frac{m_a}{|\bx - \bx_a|} + O(\epsilon^2) \,,
\nonumber \\
g_{0j} & = O(\epsilon^{3/2}) \,,
\nonumber \\
g_{jk} &= \left ( 1 + 2 \sum_a \gamma^*_a \frac{m_a}{|\bx - \bx_a|} \right ) \delta_{jk} + O(\epsilon^2) \,,
\label{eq10:EIHmetric}
\end{eqnarray}
where $\alpha^*_a$  and $\gamma^*_a$ are functions of the parameters of the theory and of the
structure of the $a$th body, and $\epsilon \sim m/r \sim v^2$.   For test-body geodesics in this metric, the quantities $\alpha^*_a m_a$, and $\sum_a \alpha^*_a m_a$ are the Kepler-measured active gravitational
masses of the individual bodies and of the system as a whole. The metric (\ref{eq10:EIHmetric}) is used mainly to discuss the propagation of photons in systems with compact bodies, and plays a role, for example,  in obtaining timing formulae for binary pulsars.  In general
relativity, $\alpha^*_a \equiv \gamma^*_a \equiv 1$.

To obtain the modified EIH Lagrangian, we first generalize the post-Newtonian
semiconservative $N$-body Lagrangian, given by Eq.\ (6.80) of \cite{tegp} or Eq.\ (6.81) of \cite{tegp2}:
\begin{eqnarray}
L & = - \sum_a m_a \left ( 1 - \frac{1}{2} v_a^2 - \frac{1}{8} v_a^4 \right )
\nonumber \\
& \quad + \frac{1}{2} \sum_a \sum_{b \ne a}  \frac{m_a m_b}{r_{ab}} \biggl [ 1 + (2\gamma + 1) v_a^2
-(2\beta-1) \sum_{c \ne a} \frac{m_c}{r_{ac}}
\nonumber \\
& \qquad 
- \frac{1}{2} (4\gamma + 3 +\alpha_1 - \alpha_2) \bv_a \cdot \bv_b 
- \frac{1}{2} (1 + \alpha_2 ) (\bv_a \cdot \bm{n}_{ab})(\bv_b \cdot \bm{n}_{ab})
\nonumber \\
& \qquad
- \xi \frac{\bx_{ab}}{r_{ab}^2} \cdot \sum_{c \ne ab} m_c \left ( \frac{\bx_{bc}}{r_{ac}}
-  \frac{\bx_{ac}}{r_{bc}}  \right )
 \biggr ] \,,
\label{eq6:lagrangian}
\end{eqnarray}
where $r_{ab} = |\bx_a - \bx_b|$ and $\bm{n}_{ab} = \bx_{ab}/r_{ab}$.   
The Lagrangian is expressed in a coordinate system at rest with respect to the preferred-frame singled out by the theory of gravity in question; this is generally assumed to coincide with the frame in which the cosmic background radiation is isotropic.  

We set the Whitehead parameter $\xi = 0$, and replace PPN parameters with parameters dependent upon each body, according to
\begin{eqnarray}
L_{\rm EIH} &= - \sum_a m_a \left [ 1 - \frac{1}{2}  v_a^2 - \frac{1}{8} \left (1+{\cal A}_a \right ) v_a^4 \right ]
\nonumber \\
& \quad  +\frac{1}{2} \sum_a  \sum_{b \ne a} \frac{m_a m_b}{r_{ab}} \left [ {\cal G}_{ab} + 3 {\cal B}_{ab} v_a^2 
- \frac{1}{2} \left ({\cal G}_{ab} + 6 {\cal B}_{(ab)} + {\cal C}_{ab} \right ) \bv_a \cdot \bv_b 
\right .
\nonumber \\
& 
\left .
\quad \quad 
- \frac{1}{2} \left ({\cal G}_{ab} + {\cal E}_{ab} \right ) (\bv_a \cdot \bm{n}_{ab})(\bv_b \cdot \bm{n}_{ab}) \right ] 
\nonumber \\
& \quad 
- \frac{1}{2} \sum_a \sum_{b \ne a}\sum_{c \ne a} {\cal D}_{abc} \frac{m_a m_b}{r_{ab}}\frac{m_c}{r_{ac}}  \,.
\label{eq10:lagrangian1}
\end{eqnarray}
The quantities
${\cal A}_a$, $ {\cal G}_{ab}$, ${\cal B}_{ab}$, ${\cal C}_{ab} $, ${\cal E}_{ab} $ and ${\cal D}_{abc}$
are functions of the parameters of the theory and of the structure of each
body, and satisfy
\begin{equation}
{\cal G}_{ab} = {\cal G}_{(ab)}  , \quad {\cal C}_{ab} = {\cal C}_{(ab)} , \quad {\cal E}_{ab} = {\cal E}_{(ab)} , \quad {\cal D}_{abc} = {\cal D}_{a(bc)} \,.
\end{equation}
Note that ${\cal B}_{ab}$ has no special symmetry, in general.

Notice that we did not introduce a parameter in front of the kinetic $v_a^2$ term in Eq.\ (\ref{eq10:lagrangian1}).  Any such parameter can always be absorbed into a new definition of the inertial mass $m'_a$ of body $a$.  We are then free to change the constant term $-\sum_a m_a$ to be the sum of the new inertial masses $-\sum_a m'_a$.  This has no effect on the equations of motion, but does allow the Hamiltonian derived from $L_{\rm EIH}$ to be the sum of the new inertial masses at lowest order.   We also did not include a term of the form
$(m_a m_b/r_{ab})(\bv_b \cdot \bm{n}_{ab})^2$; such a term can be associated (via a total time derivative in the Lagrangian) with the Whitehead term in Eq.\ (\ref{eq6:lagrangian}), which we have chosen to reject.   

In general relativity, ${\cal G}_{ab}= {\cal B}_{ab} = {\cal D}_{abc} = 1$, while ${\cal A}_a = {\cal C}_{ab} ={\cal E}_{ab} = 0$.  
In the  post-Newtonian limit of semiconservative theories (with $\xi=0$), for structureless
masses (no self-gravity), the parameters have the values [compare Eq.\ (\ref{eq6:lagrangian})]
\begin{eqnarray}
  {\cal G}_{ab}  &=1 \,, 
\quad
 {\cal B}_{ab} = \frac{1}{3} (2\gamma +1) \,, \quad {\cal D}_{abc} = 2\beta -1 \,,
\nonumber \\
{\cal A}_a &= 0 \,, \quad {\cal C}_{ab} =  \alpha_1 - \alpha_2 \,, \quad {\cal E}_{ab} =  \alpha_2 
 \,.
\end{eqnarray}
In the fully conservative case, including contributions of the 1PN-order self-gravitational binding energies of the bodies, the parameters can be shown to have the values
\begin{eqnarray}
 {\cal G}_{ab}  &= 1 + (4\beta-\gamma-3) \left ( \frac{\Omega_a}{m_a} +  \frac{\Omega_b}{m_b} \right ) \,,
\nonumber \\
 {\cal B}_{ab} &= \frac{1}{3} (2\gamma +1) \,, \quad {\cal D}_{abc} = 2\beta -1 \,,
\nonumber \\
{\cal A}_a &  = {\cal C}_{ab} =  {\cal E}_{ab} = 0 
 \,,
\end{eqnarray}
where $\Omega_a$ is the self-gravitational energy of the $a$th body.  The connection between these parameters and those introduced by Nordtvedt \cite{1985ApJ...297..390N} is detailed in an Appendix.

To obtain the Lagrangian in a moving frame, we make a Lorentz transformation from the original preferred
frame to a new frame which moves at velocity $\bm{w}$ relative to the old frame.
In order to preserve the post-Newtonian character of the Lagrangian, we
assume that $w \equiv |\bm{w}|$ is small, i.e. of $O(\epsilon^{1/2})$. This transformation from 
rest coordinates $x^\alpha=(t, \bm{x})$ to moving coordinates $\xi^\mu = (\tau, \bm{\xi})$  can be expanded in powers of
$w$ to the required order.  This approximate form of the Lorentz transformation is sometimes called a post-Galilean transformation \cite{1967RSPSA.298..123C}, and has the form
\begin{eqnarray}
\bx &= \bm{\xi} + \left ( 1+ \frac{1}{2} w^2 \right ) \bm{w} \tau + \frac{1}{2} (\bm{\xi} \cdot \bm{w} ) \bm{w} + \xi \times O(\epsilon^2) \,,
\nonumber \\
t &= \tau \left ( 1 + \frac{1}{2} w^2 + \frac{3}{8} w^4 \right ) + \left ( 1+ \frac{1}{2} w^2 \right ) \bm{\xi} \cdot \bm{w} + \tau \times O(\epsilon^3) \,,
\label{eq10:lorentz}
\end{eqnarray}
where $\bm{w} \tau$ is assumed to be $O(\epsilon^0)$.  
From the transformation (\ref{eq10:lorentz}), we have
\begin{eqnarray}
\bv_a &= \bm{\nu}_a + \bm{w} - \frac{d\bm{\nu}_a}{d\tau} (\bm{\xi}_a \cdot \bm{w}) - \frac{1}{2} w^2 \bm{\nu}_a - (\bm{\nu}_a \cdot \bm{w}) \left (\bm{\nu}_a + \frac{1}{2} \bm{w} \right ) \,,
\nonumber \\
\frac{1}{r_{ab}} &= \frac{1}{\xi_{ab}} \left [ 1 + \frac{1}{2} (\bm{w} \cdot \bm{n}'_{ab} )^2 
+ \frac{1}{\xi_{ab}}  (\bm{w} \cdot \bm{\xi}_a )(\bm{\nu}_a \cdot \bm{n}'_{ab} )
\right .
\nonumber \\
& \qquad \left .
-  \frac{1}{\xi_{ab}}(\bm{w} \cdot \bm{\xi}_b )(\bm{\nu}_b \cdot \bm{n}'_{ab} ) \right ] \,,
\end{eqnarray}
where $\bm{\xi}_a$, $\bm{\xi}_b$ and  $\bm{\nu}_a \equiv d\bm{\xi}_a/d\tau$ are to be evaluated at the same time $\tau$, given by a clock at the spatial origin ($\bm{\xi}=0$) of the moving coordinate system, and $\bm{n}'_{ab} \equiv \bm{\xi}_{ab}/{\xi}_{ab}$.  
The new Lagrangian is given by
\begin{equation}
L( \bm{\xi}, \tau) = L(\bx, t) \frac{dt}{d\tau} - \frac{df}{d\tau} \,,
\label{eq10:lagrangian2}
\end{equation}
where $dt/d\tau$ is evaluated at $\bm{\xi}=0$, and where we are free to subtract a total time derivative of a function $f$ to simplify the new Lagrangian.
Substituting these results into Eqs.\ (\ref{eq10:lagrangian1}) and (\ref{eq10:lagrangian2}), 
dropping constants and total time derivatives, and replacing $\xi_{ab}$ and $\bm{\nu}_a$ with $r_{ab}$ and $\bv_a$, we obtain the Lagrangian in the moving frame
\begin{eqnarray}
L&=  L_{\rm EIH}  + \frac{1}{4} \sum_a m_a {\cal A}_a \left [ v_a^2 w^2 + 2v_a^2 (\bm{v}_a \cdot \bm{w}) + 2(\bm{v}_a \cdot \bm{w})^2 \right ]
\nonumber \\
& \quad - \frac{1}{4} \sum_{a \ne b} \frac{m_a m_b}{r_{ab}}
 \left \{  {\cal C}_{ab} w^2 - 2\left ( 6 {\cal B}_{[ab]} - {\cal C}_{ab} \right ) (\bm{v}_a \cdot \bm{w})
 \right .
 \nonumber \\
 &
 \quad \quad 
 \left. 
 + {\cal E}_{ab} \left [ (\bm{w} \cdot \bm{n}_{ab} )^2 + 2 (\bm{w} \cdot \bm{n}_{ab} )(\bm{v}_a \cdot \bm{n}_{ab} ) \right ]
 \right \} 
 \,.
\label{eq10:lagrangian3}
\end{eqnarray}
Notice that the Lagrangian is post-Galilean invariant if and only if
\begin{equation}
{\cal A}_a  \equiv {\cal B}_{[ab]} \equiv {\cal C}_{ab} \equiv {\cal E}_{ab} \equiv 0 \,.
\label{eq10:postGalconditions}
\end{equation}
These quantities are then the preferred-frame parameters of our modified EIH formalism.

\section{Modified EIH parameters in alternative theories}
\label{sec:alternative}

As an illustration of this modified EIH framework for compact bodies, we will focus on specific theories where calculations have been carried out.   As we have already discussed, a variety of approaches have shown that  the EIH equations of motion for compact objects within general relativity are identical to those of the post-Newtonian limit with weak fields everywhere.  In other words, in general relativity, ${\cal G}_{ab} \equiv {\cal B}_{ab} \equiv {\cal D}_{abc} \equiv 1$, and the remaining coefficients vanish, independently of
the nature of the bodies.

\subsection{Scalar-tensor theories}
\label{sec10:STmotion}

The modified EIH formalism was first developed by Eardley \cite{1975ApJ...196L..59E} for application to the Jordan-Fierz-Brans-Dicke theory.  It makes use of the fact that
only the scalar fleld $\phi$ produces an external influence on the structure of
each compact body via its boundary values in the matching region. This boundary value of $\phi$ is related to the local value of the gravitational
constant as felt by the compact body by
\begin{equation}
G_{\rm local} = \frac{G}{\phi} \left (\frac{4 + 2\omega}{3+2\omega} \right ) \,,
\label{eq10:glocal}
\end{equation}
where $G$ is the fundamental gravitational coupling constant.   Thus we will treat the inertial mass $m_a$ of each body as a being a function of $\phi$.   Then, by defining the deviation of $\phi$ from its asymptotic value $\phi_0$ by $\phi \equiv \phi_0 (1+ \Psi)$, we can write down the expansion 
\begin{equation}
m_a(\phi) =
 m_{a}\left [1  +  s_a \Psi + \frac{1}{2}  (s_a^2 +s'_a   -s_a) \Psi^2 + O(\Psi^3)  \right ]\,,
 \label{eq10:mexpand2}
\end{equation}  
where $m_a \equiv m_a (\phi_0)$, and  we define the dimensionless sensitivities
\begin{eqnarray}
s_a &\equiv \left ( \frac{d \ln m_a(\phi)}{d \ln \phi} \right )_0 \,,
\nonumber \\
s'_a &\equiv \left ( \frac{d^2 \ln m_a(\phi)}{d (\ln \phi)^2} \right )_0 \,.
\label{eq10:sensitivitiesST}
\end{eqnarray}
The action for massless scalar-tensor theory is then written
\begin{equation}
I = \frac{1}{16\pi G} \int \left [ \phi R - \frac{\omega(\phi)}{\phi} g^{\mu\nu} \phi_{,\mu} \phi_{,\nu}  \right ] \sqrt{-g} d^4x  - \sum_a \int m_a (\phi) d\tau_a \,,
\label{eq10:STaction1}
\end{equation}
where the integrals over proper time  $\tau_a$ are to be taken along the world line of each body $a$.  It is straightforward to vary the action with respect to $g_{\mu\nu}$ and $\phi$ to obtain the field equations,
\begin{eqnarray}
G_{\mu\nu} &= \frac{8\pi G}{\phi} T_{\mu\nu} + \frac{\omega(\phi)}{\phi^2} \left ( \phi_{,\mu} \phi_{,\nu} - \frac{1}{2} g_{\mu\nu} \phi_{,\lambda} \phi^{,\lambda} \right ) 
\nonumber \\
& \quad+ \frac{1}{\phi} \left ( \phi_{;\mu\nu} - g_{\mu\nu} \Box_g \phi \right ) \,, 
\label{eq10:STfieldeq1}\\
\Box_g \phi  &= \frac{1}{3 + 2\omega(\phi)} \left ( 8\pi G T  - 16\pi G \phi  \frac{\partial T}{\partial \phi} - \frac{d\omega}{d\phi} \phi_{,\lambda} \phi^{,\lambda}  \right ) \,,
\label{eq10:STfieldeq2}
\end{eqnarray}
\label{eq10:STfieldeq}
where 
\begin{equation}
T^{\mu\nu} = (-g)^{-1/2} \sum_a m_a (\phi) u_a^\mu u_a^\nu (u_a^0)^{-1} \delta^3 ({\bf x} - {\bf x}_a)  \,,
\end{equation}
where $u_a^\mu$ is the four-velocity of body $a$.
The equations of motion take the form
\begin{equation}
{T^{\mu\nu}}_{;\nu} - \frac{\partial T}{\partial \phi} \phi^{,\nu} = 0 \,.
\end{equation}

Carrying out a post-Newtonian calculation of the metric as described, for example in \cite{2013PhRvD..87h4070M} or in Sec.\ 5.3 of \cite{tegp2}, we obtain, to lowest order 
\begin{eqnarray}
\Psi & =2 \zeta \sum_b \frac{m_b}{r_b} \left (1 - 2s_b \right ) + O(\epsilon^2 ) \,,
\nonumber \\
g_{00} & = -1 + 2 \sum_b \frac{m_b}{r_b} \left ( 1 - 2 \zeta s_b \right ) + O(\epsilon^2 ) \,,
\nonumber \\
g_{0j} &= - 4 (1-\zeta) \sum_b \frac{m_b v_b^j}{r_b} + O(\epsilon^{5/2}) \,,
\nonumber \\
g_{jk} &= \delta_{jk} \left [ 1 + 2 \sum_b \frac{m_b}{r_b} \left ( 1 - 2 \zeta +2 \zeta s_b \right ) \right ] \,,
\label{eq10:Psimetric}
\end{eqnarray}
where $\zeta = 1/(4 + 2\omega_0)$, $r_b = |\bx-\bx_b|$, and we have chosen units in which $G_{\rm local} =1$.  For the explicit $O(\epsilon^2)$ terms in $g_{00}$ and $\Psi$, see \cite{2013PhRvD..87h4070M}.  Notice that the active gravitational mass as measured by test-body Keplerian orbits far from each body is given by 
\begin{equation}
(m_{\rm A})_a = m_a (1- 2 \zeta s_a ) \,.
\label{eq10:mactive1}
\end{equation}

From the complete post-Newtonian solution for $g_{\mu\nu}$ and $\Psi$, we can obtain the matter  action for the $a$th body, given by 
\begin{equation}
I_a = - \int m_a(\phi) \left ( - g_{00} - 2 g_{0j} v_a^j - g_{jk} v_a^j v_a^k \right )^{1/2} dt \,.
\end{equation}
To obtain an $N$-body action in the form of Eq.\ (\ref{eq10:EIH}), we first make
the gravitational terms in $I_a$, manifestly symmetric under interchange of
all pairs of particles, then take one of each such term generated in $I_a$, and
sum the result over $a$.  The resulting $N$-body Lagrangian then has the form of Eq.\ (\ref{eq10:lagrangian1}) with
\begin{eqnarray}
{\cal G}_{ab} &= 1 - 2\zeta \left (s_a +s_b  - 2s_a s_b \right ) \,,
\nonumber\\
{\cal B}_{ab} &= \frac{1}{3} \left [ {\cal G}_{ab} + 2(1-\zeta) \right ] \,,
\nonumber \\
{\cal D}_{abc} &= {\cal G}_{ab} {\cal G}_{ac} + 2\zeta (1-2s_b)(1-2s_c) \left [ \lambda (1-2s_a) + 2 \zeta s'_a \right ] \,,
\nonumber \\
{\cal A}_a &= {\cal B}_{[ab]} =  {\cal C}_{ab} = {\cal E}_{ab} = 0 \,.
\label{eq10:EIHcoefficients}
\end{eqnarray}

\subsection{Vector-tensor theories}
\label{sec10:VTmotion}

Because the norm of the vector field $\bm{K}$ in Einstein-{\AE}ther and Khronometric theories is constrained to be $-1$,
the structure of a spherically symmetric compact body at rest with rest to the preferred rest frame does not depend on it.  However it could depend on the time component $K^0$, or more properly on the invariant quantity ${\bm K} \cdot {\bm u}$, where $\bm u$ is the body's four-velocity.  (The structure of a  rotating body could also depend on the projection of $\bm K$ along the body's spin axis, but here we will focus on non-rotating bodies.)   We define for a body with four-velocity $\bm u$, 
\begin{equation}
\gamma \equiv - K^\mu  u_\mu \equiv 1 + \Psi  \,,
\end{equation}
where we assume that far from the system, for a test body at rest, ${\bm K} \cdot {\bm u} = -1$.    We define the sensitivities 
\begin{eqnarray}
s_a &\equiv \left ( \frac{d \ln m_a (\gamma)}{d \ln \gamma} \right )_{\gamma =1} \,,
\nonumber \\
s'_a &\equiv \left ( \frac{d^2 \ln m_a (\gamma)}{d (\ln \gamma)^2 } \right )_{\gamma =1} \,,
\end{eqnarray}  
where the derivatives are to be taken holding baryon number fixed.  Then the expansion of $m_a(\gamma)$ in powers of $\Psi$ is again given by Eq. (\ref{eq10:mexpand2}).
With this assumption, Foster \cite{2007PhRvD..76h4033F} and Yagi et al. \cite{2014PhRvD..89h4067Y}  derived the metric and equations of motion to post-Newtonian order for systems of compact bodies in Einstein-{\AE}ther theory.  In the preferred rest frame, the metric is given by
\begin{eqnarray}
g_{00} &= -1 + 2U - 2U^2 - 2\Phi_2 + 3\Phi_{1s} + O(\epsilon^3) \,,
\nonumber \\
g_{0j} &= g^j + O(\epsilon^{5/2}) \,,
\nonumber \\
g_{jk} &= (1+2U) \delta_{jk} + O(\epsilon^2) \,,
\label{eq10:AEmetric}
\end{eqnarray}
and the vector field is given by
\begin{eqnarray}
K^0 &= (-g_{00})^{-1/2} + O(\epsilon^3) \,,
\nonumber \\
K^j &= k^j + O(\epsilon^{5/2}) \,, 
\label{eq10:AEvector}
\end{eqnarray}
where
\begin{eqnarray}
U & = \sum_b \frac{G_N m_b}{r_b} \,, \quad \Phi_2 = \sum_{b,c} \frac{G_N^2 m_b m_c}{r_b r_{bc}}  \,, \quad \Phi_{1s} = \sum_b \frac{G_N m_b}{r_b} v_b^2 (1-s_b) \,,
\nonumber \\
g^j &= \sum_b  \frac{G_N m_b}{r_b} \left [ B_b^{-} v_b^j + B_b^{+} n_b^j (\bm{n}_b \cdot \bv_b) \right ] \,,
\nonumber \\
k^j &= \sum_b  \frac{G_N m_b}{r_b} \left [ C_b^{-} v_b^j + C_b^{+} n_b^j (\bm{n}_b \cdot \bv_b) \right ] \,,
\end{eqnarray}
where $G_N = 2G/(2-c_{14})$, $G$ is the gravitational coupling constant of the theory,  $\bm{n}_b = (\bm{x} - \bm{x}_b)/r_b$, and $c_{14} = c_1 +c_4$.   
The quantities $B_b^{\pm}$ and $C_b^{\pm}$ are complicated expressions involving the constants $c_1$, $c_2$, $c_3$ and $c_4$ of Einstein-{\AE}ther theory and the sensitivities $s_b$ (see Eqs.\ (23) and (29) of \cite{2007PhRvD..76h4033F}).   

From Eqs.\ (\ref{eq10:AEmetric}) and (\ref{eq10:AEvector}) we obtain
\begin{equation}
\Psi (\bx_a)  = \frac{1}{2} v_a^2 + \frac{3}{8} v_a^4 + 2 v_a^2 U(\bx_a) - v_a^j k^j(\bx_a)  + O(\epsilon^3) \,. 
\label{eq10:Psiexpand}
\end{equation}
Note that, when $v_a^j =0$, $\Psi = 0$, resulting in no dependence of the inertial mass on the vector field.  
Writing the action for the $a$th body as 
\begin{equation}
I_a = - \int m_a(\Psi) \left ( - g_{00} - 2 g_{0j} v_a^j - g_{jk} v_a^j v_a^k \right )^{1/2} dt \,,
\end{equation}
we expand to post-Newtonian order, make the action manifestly symmetric under interchange of all pairs of particles, select one of each term, and sum over $a$.   After rescaling each mass by $m_a \to m_a/(1-s_a)$ and replacing the constant term in the Lagrangian by the sum of the rescaled masses, we obtain the modified EIH Lagrangian in the form  of Eq.\ (\ref{eq10:lagrangian1}), with
\begin{eqnarray}
{\cal G}_{ab} &= \frac{G_N}{(1-s_a)(1-s_b)} \,, 
 \nonumber \\
 {\cal B}_{ab} &= {\cal G}_{ab} (1-s_a) \,,
\nonumber \\
{\cal D}_{abc} &={\cal G}_{ab} {\cal G}_{ac} (1-s_a) \,,
\nonumber \\
 {\cal A}_a &=  s_a - \frac{s'_a}{1-s_a} \,,
\nonumber \\
{\cal C}_{ab} &=  {\cal G}_{ab} \left [  \alpha_1 - \alpha_2 + 3 \left (s_a + s_b \right )  - {\cal Q}_{ab} -{\cal R}_{ab}  \right ] \,,
\nonumber \\
{\cal E}_{ab} &= {\cal G}_{ab} \left [  \alpha_2 + {\cal Q}_{ab} - {\cal R}_{ab}  \right ] \,,
\label{eq10:EIHcoeffs1}
\end{eqnarray}
where $\alpha_1$ and $\alpha_2$ are the PPN preferred-frame parameters of Einstein-{\AE}ther theory, given by 
\begin{eqnarray}
  \alpha_1 & =  -\frac{8 (c_3^2 + c_1 c_4)}{2 c_1 - c_1^2 + c_3^2} \,,
\nonumber  \\
  \alpha_2 & =  \frac{1}{2} \alpha_1 - \frac{(2c_{+}-c_{14})(c_{+}+c_{14}+3c_2)}{c_{123} (2-c_{14})}\,,
   \label{eq5:PPNAE}
\end{eqnarray}
and 
\begin{eqnarray}
{\cal Q}_{ab} &= \frac{1}{2} \left (\frac{2-c_{14}}{2c_{+} - c_{14}} \right ) (\alpha_1 - 2\alpha_2) (s_a +s_b) +  \frac{2 - c_{14}}{c_{123}} s_a s_b  \,,
\nonumber \\
{\cal R}_{ab} &= \frac{8+\alpha_1}{4c_1} \left [ c_{-} (s_a + s_b) + (1-c_{-}) s_as_b \right ]\,.
\label{eq10:EIHcoeffs2}
\end{eqnarray}
Here 
 $c_\pm = c_1 \pm c_3$, and $c_{123} = c_1 + c_2 +c_3$.   The two-body equations of motion that follow from the Lagrangian with these coefficients agree with Eq.\ (33) of \cite{2007PhRvD..76h4033F} (after correcting a sign and a parenthesis in Eqs.\ (34) and (35) of that paper).

Note that the Lagrangian is in general not Lorentz invariant, and therefore will exhibit preferred-frame effects.  Even when the parameters $c_i$ are constrained so as to enforce $\alpha_1 = \alpha_2 = 0$, making the dynamics Lorentz invariant in the fully weak field post-Newtonian limit, the dynamics of compact bodies can still be dependent on the overall motion of the system via the motion-induced sensitivities of the bodies.

The corresponding equations in Khronometric theory can be obtained from these by setting $c_1 = - \epsilon$, $c_2 = \lambda_K$, $c_3 = \beta_K + \epsilon$, and $c_4 = \alpha_K + \epsilon$, and taking the limit $\epsilon \to \infty$ \cite{2014PhRvD..89h4067Y}.  The parameters of the modified EIH Lagrangian are given by Eqs.\ (\ref{eq10:EIHcoeffs1}), but now with $G_N = 2G/(2-\alpha_K)$, and 
\begin{eqnarray}
{\cal Q}_{ab} &=  \frac{1}{\beta_K + \lambda_K} \left [(\alpha_K + \beta_K + 3 \lambda_K) (s_a +s_b) + (2-\alpha_K)  s_a s_b \right ] \,,
\nonumber \\
{\cal R}_{ab} &= \frac{1}{2} (8+\alpha_1) \left [ s_a + s_b - s_as_b \right ]\,,
\label{eq10:EIHcoeffs3}
\end{eqnarray}
where $\alpha_1$ is the PPN parameter of Khronometric theory, given by 
\begin{equation}
\alpha_1 = \frac{4(\alpha_K - 2 \beta_K)}{\beta_K -1} \,.
\end{equation}

Yagi et al. \cite{2014PhRvD..89h4067Y}  calculated neutron star sensitivities in both Einstein-{\AE}ther  and Khronometric theories.  In order to do so, it was necessary to construct models for neutron stars moving uniformly relative to the preferred frame.  From
Eqs.\ (\ref{eq10:mexpand2}) and (\ref{eq10:Psiexpand}), assuming uniform motion with no external bodies ($U = k^j =0$), the sensitivity is given by $s =  v^{-1} d \ln m/dv$.   They chose the coefficients $c_2$ and $c_4$ so that the PPN parameters $\alpha_1$ and $\alpha_2$ saturate the bounds from solar-system measurements, and obtained fitting formulae for the sensitivities as a function of $c_+$, $c_{-}$ and the compactness $M/R$ of the neutron star.

\section{Application to two- and three-body dynamics}
\label{sec:applications}

Since one of our goals is to apply this formalism to binary systems
containing compact objects, let us now restrict
attention to two-body systems.  
We obtain from $L$ the two-body equations of motion
\begin{eqnarray}
\bm{a}_1 &=  -\frac{m_2 \bm{n}}{r^2} \left \{{\cal G}_{12} 
- \left (3{\cal G}_{12}{\cal B}_{12} + {\cal D}_{122} \right ) \frac{m_2}{r}
\right .
\nonumber \\
& 
\left .
\quad \quad
- \frac{1}{2} \left [ 2{\cal G}_{12}^2 + 6{\cal G}_{12}{\cal B}_{(12)} + 2{\cal D}_{211} + {\cal G}_{12}({\cal C}_{12} + {\cal E}_{12}) \right ] \frac{m_1}{r} 
\right .
\nonumber \\
& 
\left .
\quad \quad
+ \frac{1}{2}  \left [3{\cal B}_{12} - {\cal G}_{12} (1+ {\cal A}_1) \right ] v_1^2
+ \frac{1}{2}(3{\cal B}_{21} + {\cal G}_{12}+ {\cal E}_{12}) v_2^2 
\right .
\nonumber \\
& 
\left .
\quad \quad
- \frac{1}{2} \left ( 6{\cal B}_{(12)} + 2 {\cal G}_{12} + {\cal C}_{12} + {\cal E}_{12} \right ) \bm{v}_1 \cdot \bm{v}_2  
- \frac{3}{2} \left ({\cal G}_{12}+ {\cal E}_{12} \right ) (\bm{n} \cdot \bv_2)^2 
\right .
\nonumber \\
& 
\left .
\quad \quad
+ \frac{1}{2} \left ( {\cal C}_{12} + {\cal G}_{12}  {\cal A}_1 \right ) w^2
+ \frac{1}{2} \left ( {\cal C}_{12} - 6{\cal B}_{[12]}+ {\cal E}_{12} + 2 {\cal G}_{12}  {\cal A}_1 \right ) \bv_1 \cdot \bm{w}
\right .
\nonumber \\
& 
\left .
\quad \quad
+ \frac{1}{2} \left ( {\cal C}_{12} + 6{\cal B}_{[12]}- {\cal E}_{12}  \right ) \bv_2 \cdot \bm{w}
\right .
\nonumber \\
& 
\left .
\quad \quad
+ \frac{3}{2} {\cal E}_{12} \left [ (\bm{w} \cdot \bm{n} )^2 + 2(\bm{w} \cdot \bm{n} )(\bm{v}_2 \cdot \bm{n} ) \right ]
\right \}
\nonumber \\
& \quad
+  \frac{m_2 \bm{v}_1}{r^2} \bm{n} \cdot \left \{ \left [3{\cal B}_{12} + {\cal G}_{12} (1+ {\cal A}_1) \right ] \bv_1 - 3{\cal B}_{12} \bv_2  + {\cal G}_{12}  {\cal A}_1 \bm{w}  \right \} 
\nonumber \\
& \quad
- \frac{1}{2} \frac{m_2 \bm{v}_2}{r^2} \bm{n} \cdot \left \{
\left ( 6{\cal B}_{(12)} + 2 {\cal G}_{12} + {\cal C}_{12} + {\cal E}_{12} \right ) \bv_1
\right .
\nonumber \\
& 
\left .
\quad \quad
-\left ( 6{\cal B}_{(12)}  + {\cal C}_{12} - {\cal E}_{12} \right ) \bv_2
+2 {\cal E}_{12} \bm{w} 
 \right \} 
 \nonumber \\
& \quad
- \frac{1}{2} \frac{m_2 \bm{w}}{r^2} \bm{n} \cdot \left \{
\left (  {\cal C}_{12} -6{\cal B}_{[12]} + {\cal E}_{12} -2{\cal G}_{12}  {\cal A}_1 \right ) \bv_1
\right .
\nonumber \\
& 
\left .
\quad \quad
- \left ({\cal C}_{12} - 6{\cal B}_{[12]} - {\cal E}_{12}  \right ) \bv_2
-2 \left ( {\cal G}_{12}  {\cal A}_1 - {\cal E}_{12}  \right ) \bm{w} 
 \right \} 
\,,
\nonumber \\
\bm{a}_2 &= \{ 1\rightleftharpoons 2; \, \bm{n} \to -\bm{n} \} \,,
\label{eq10:2bodyeom}
\end{eqnarray}
where $\bm{a}_a \equiv d\bv_a/dt$, $\bx \equiv \bx_1 - \bx_2$, $r \equiv |\bx|$, and $\bm{n} \equiv \bx/r$.  If the conditions of Eq. (\ref{eq10:postGalconditions}) hold, then the preferred-frame terms vanish.

To Newtonian order, the center of mass $\bm{X} = (m_1 \bx_1 + m_2 \bx_2)/m$ of the system is unaccelerated, thus, to sufficient accuracy in the post-Newtonian terms, we can set $\bm{X} = \dot{\bm{X}} = 0$ and 
write
\begin{eqnarray}
\bx_1 &= \left [ \frac{m_2}{m} + O(\epsilon) \right ] \bx \,,
\nonumber \\
\bx_2 &= -\left [ \frac{m_1}{m} + O(\epsilon) \right ] \bx \,.
\end{eqnarray}
We define
\begin{eqnarray}
\bv &\equiv \bv_1 - \bv_2 \,, \quad \bm{a} \equiv \bm{a}_1 - \bm{a}_2 \,,
\nonumber\\
m &\equiv m_1 +m_2 \,, \quad \eta \equiv \frac{m_1m_2}{m^2} \,, \quad \Delta = \frac{m_2 - m_1}{m} \,.
\end{eqnarray}
We also define the two-body coefficients
\begin{eqnarray}
{\cal G} &\equiv {\cal G}_{12} \,, \quad {\cal B}_{+} \equiv {\cal B}_{(12)} \,, \quad {\cal B}_{-} \equiv {\cal B}_{[12]} \,, \quad 
{\cal D} \equiv \frac{m_2}{m} {\cal D}_{122} + \frac{m_1}{m} {\cal D}_{211} \,, 
\nonumber \\
{\cal C} &\equiv {\cal C}_{12} \,, \quad {\cal E} \equiv {\cal E}_{12} \,, \quad
{\cal A}^{(n)} \equiv \left ( \frac{m_2}{m} \right )^n {\cal A}_1 - \left (- \frac{m_1}{m} \right )^n {\cal A}_2  \,.
\end{eqnarray}
Then the equation of motion for the relative orbit takes the form 
\begin{equation}
\bm{a} = \bm{a}_{\rm L} + \bm{a}_{\rm PF} \,,
\end{equation}
where the purely two-body, or ``local'' contributions have the form (we use ``hats'' to denote parameters associated with compact bodies)
\begin{equation}
\bm{a}_{\rm L} =  -\frac{{\cal G} m \bm{n}}{r^2} 
+ \frac{m}{r^2} \left [ \bm{n} \left ( \hat{A}_1 v^2 + \hat{A}_2 \rdot^2 + \hat{A}_3 \frac{m}{r} \right ) 
+ \rdot \bv \hat{B}_1 \right ] \,,
\label{eq10:relativeacc1}
\end{equation}
where
\begin{eqnarray}
\hat{A}_1 &= \frac{1}{2} \left \{ {\cal G} \left ( 1- 6\eta \right ) - 3{\cal B}_{+} - 3 \Delta {\cal B}_{-} - \eta ({\cal C} + 2{\cal E}) +  {\cal G} A^{(3)}   \right \} \,,
\nonumber \\
\hat{A}_2 &= \frac{3}{2} \eta ({\cal G} + {\cal E}) \,,
\nonumber \\
\hat{A}_3 &= {\cal G} \left [ 2 \eta {\cal G} + 3{\cal B}_{+} + \eta \left ( {\cal C} + {\cal E} \right ) + 3 \Delta {\cal B}_{-} \right ] + {\cal D} \,,
\nonumber \\
\hat{B}_1 &= {\cal G}  (1-2\eta)  + 3{\cal B}_{+} + 3 \Delta {\cal B}_{-} + \eta  {\cal C} +  {\cal G} A^{(3)} \,,
\label{eq10:relativeacc2}
\end{eqnarray}
and the preferred-frame contributions have the form
\begin{eqnarray}
\bm{a}_{\rm PF} &=  
 \frac{m}{r^2} \left \{ \bm{n} \left [ \left ( \frac{1}{2} \hat{\alpha}_1 + 2{\cal G} {\cal A}^{(2)} \right ) (\bm{w} \cdot \bv ) + \frac{3}{2} \left (\hat{\alpha}_2+ {\cal G} {\cal A}^{(1)} \right ) (\bm{w} \cdot \bm{n})^2 \right ]
\right .
\nonumber \\
&
\quad
\left . 
- \bm{w} \left [ \frac{1}{2} \hat{\alpha}_1 (\bm{n} \cdot \bv )
+ \hat{\alpha}_2 (\bm{n} \cdot \bm{w}) \right ]
+ {\cal G} {\cal A}^{(2)}  \bv (\bm{n} \cdot \bm{w})
\right \}  \,,
\label{eq10:relativeacc3}
\end{eqnarray}
where 
\begin{eqnarray}
\hat{\alpha}_1 & = \Delta \left ( {\cal C} + {\cal E}  \right)
  - 6 {\cal B}_{-}  - 2 {\cal G} {\cal A}^{(2)} \,,
  \nonumber \\
\hat{\alpha}_2 & = {\cal E} - {\cal G} {\cal A}^{(1)} 
 \,.
\label{eq10:relativeacc4}
\end{eqnarray}
These two parameters play the role of compact-body analogues of the PPN parameters $\alpha_1$ and $\alpha_2$ in the context of binary systems.

In the Newtonian limit of the orbital motion, we have $\bm{a} = {\cal G}m\bx/r^3$, with Keplerian orbit solutions given by ${\bm x}= r {\bm n}$, $r=p/(1+e \cos f)$, where
\begin{eqnarray}
{\bm n} &\equiv \left ( \cos \Omega \cos \phi- \cos \iota \sin \Omega \sin \phi \right ) {\bm e}_X 
\nonumber \\
& \quad
 + \left ( \sin \Omega \cos \phi + \cos \iota \cos \Omega \sin \phi \right ){\bm e}_Y
+ \sin \iota \sin \phi {\bm e}_Z \,,
\label{eq6:keplerorbit}
\end{eqnarray}
where the orbit elements are: semilatus rectum $p$, eccentricity $e$, inclination $\iota$, longitude of ascending node $\Omega$ and pericenter angle $\omega$; 
$f \equiv \phi - \omega$ is the  true anomaly, $\phi$ is the orbital phase measured from the ascending node and 
${\bm e}_A$ are chosen reference basis vectors.   We also have that  ${\bm v} = \dot{r} {\bm n} + (h/r) {\bm \lambda}$ and $\dot{r} = (he/p) \sin f$, where $\bm{\lambda} = \partial \bm{n}/\partial \phi$.  The orbital angular momentum per unit reduced mass is given by $\bm{h} = \bm{x} \times \bm{v} = ({\cal G}mp)^{1/2} \hat{\bm{h}}$, where $\hat{\bm{h}} = \bm{n} \times \bm{\lambda}$.   The semilatus rectum $p$ is related to the semimajor axis $a$ by $p = a(1-e^2)$.  

In the presence of perturbations, these equations define the ``osculating'' orbit, with variable orbital elements.   
Folowing the standard methods for the perturbed Kepler problem as described, for example, in Sec.\ 3.3 of \cite{2014grav.book.....P}, we obtain the secular changes in the orbit elements. 
From the post-Newtonian terms in Eq.\ (\ref{eq10:relativeacc1}), we find that the pericenter advance per orbit is given by
\begin{equation}
\Delta \omega = \frac{6 \pi m}{p} {\cal P} {\cal G}^{-1} \,,
\label{eq10:deltaomega}
\end{equation}
where 
\begin{eqnarray}
{\cal P} &= {\cal G}{\cal B}_{+} + \frac{1}{6} \left ({\cal G}^2 -   {\cal D} \right )
+ \frac{1}{6} {\cal G} \left [  6  \Delta {\cal B}_{-}  +\eta {\cal G} (2{\cal C} + {\cal E} )  + {\cal G}{\cal A}^{(3)}  \right ]\,.
\label{eq10:deltaomega2}
\end{eqnarray}
 This is the only secular perturbation produced by the post-Newtonian terms in Eq.\ (\ref{eq10:relativeacc1}).  
 
We now calculate the secular changes in the orbit elements resulting from the preferred-frame perturbations in Eq.\ (\ref{eq10:relativeacc3}).  We define
\begin{eqnarray}
\bm{e}_{\rm P} &\equiv \bm{n}|_{\phi = \omega} = \bm{e}_\Omega \cos \omega + \bm{e}_\perp \sin \omega \,,
\nonumber \\
\bm{e}_{\rm Q} &\equiv \bm{\lambda}|_{\phi = \omega} = - \bm{e}_\Omega \sin \omega + \bm{e}_\perp \cos \omega \,,
\nonumber \\
\hat{\bm h} &\equiv \bm{e}_{\rm P} \times \bm{e}_{\rm Q} =  \bm{e}_{\Omega} \times \bm{e}_{\perp}\,,
\label{eq8:basisvectors}
\end{eqnarray}
where $\bm{e}_{\rm P}$ is a unit vector pointing toward the pericenter and $\bm{e}_{\rm Q}=\hat{\bm h}  \times \bm{e}_{\rm P}$; $\bm{e}_\Omega$ is a unit vector pointing along the ascending node, and $\bm{e}_{\perp} = \hat{\bm h} \times \bm{e}_{\Omega}$.   For any vector $\bm A$, we then define components $A_{\rm P}$, $A_{\rm Q}$, $A_{\rm h}$, $A_\Omega$ and $A_\perp$ accordingly.  

The secular changes in the orbit elements are then given by
\begin{eqnarray}
 \Delta a &= 0 \,,
\nonumber \\
 \Delta e &= -\pi \hat{\alpha}_1 \Delta \left (\frac{m}{p} \right )^{1/2} w_{\rm P} (1-e^2) F(e) 
+2\pi \hat{\alpha}_2  w_{\rm P}w_{\rm Q} e\sqrt{1-e^2} F(e)^2 
 \,,
\nonumber \\
\Delta \varpi &= - \pi \hat{\alpha}_1 \Delta \left (\frac{m}{p} \right )^{1/2} w_{\rm Q} \frac{\sqrt{1-e^2} F(e)}{e} 
-\pi \hat{\alpha}_2  \left (w_{\rm P}^2 - w_{\rm Q}^2 \right )  F(e)^2 
 \,,
\nonumber \\
 \Delta \iota &=  \pi \hat{\alpha}_1 \Delta \left (\frac{m}{p} \right )^{1/2} w_{\rm h} \sin(\omega) e F(e)
-2\pi \hat{\alpha}_2 w_{\rm h} w_{\rm R} \frac{F(e)}{\sqrt{1-e^2}}
 \,,
\nonumber \\
 \Delta \Omega &= - \pi \hat{\alpha}_1 \Delta \left (\frac{m}{p} \right )^{1/2} \frac{w_{\rm h}}{\sin \iota} \cos(\omega) e F(e)
-2\pi \hat{\alpha}_2 \frac{w_{\rm h} w_{\rm S}}{\sin \iota} \frac{F(e)}{\sqrt{1-e^2}}
 \,,
 \label{eq8:deltaelements1}
\end{eqnarray}
where $\Delta \varpi = \Delta \omega + \cos \iota \, \Delta \Omega$, 
$F(e) \equiv (1+ \sqrt{1-e^2})^{-1}$,
and for any vector $\bm{A}$,
\begin{eqnarray}
A_{\rm R} &\equiv A_{\rm P} \cos(\omega) - A_{\rm Q} \sin(\omega) \sqrt{1-e^2} \,,
\nonumber \\
A_{\rm S} &\equiv A_{\rm P} \sin(\omega) + A_{\rm Q} \cos(\omega) \sqrt{1-e^2} \,.
\label{eq8:ARAS}
\end{eqnarray}
Note that the secular perturbations do not depend on the coefficients ${\cal G}{\cal A}^{(n)}$.

It is tempting to interpret these secular changes as implying linearly growing values of the orbital elements.  However, the expressions in Eqs.\ (\ref{eq8:deltaelements1}) depend on $\omega$, both from explicit $\omega$ dependence, and via the P, Q, R and S components of $\bm{w}$.  The pericenter angle is actually advancing at an average rate $d\omega/d\phi \equiv \omega' \simeq 3(m/p) {\cal P} {\cal G}^{-1}$ [see Eq.\ (\ref{eq10:deltaomega})], which we anticipate is much larger than the preferred-frame effects shown in Eq.\ (\ref{eq8:deltaelements1}) -- the goal is to set strong upper bounds on such effects.   Thus the variations in the orbit elements will be modulated on a pericenter precession timescale and could even change sign.    So in order to find the proper long-term evolution of the elements, we define, for a given element $X_a$, $dX_a/d\phi \equiv \Delta X_a /2\pi$, insert $\omega = \omega_0 + \omega' \phi$ in the right-hand-sides of Eqs.\ (\ref{eq8:deltaelements1}), including in the P, Q, R and S components of various vectors, and integrate with respect to $\phi$.   As we will be interested in low-eccentricity binary pulsars, we will assume that $e \ll 1$.   Inserting the resulting integrals for $e$ and  $\varpi$, into the expression
\begin{equation}
r = a [1-e\cos(\phi - \omega) + O(e^2)] \,,
\label{eq8:requation}
\end{equation}
and expanding to first order in the preferred-frame perturbations, we obtain
\begin{eqnarray}
\frac{r}{a} &= 1 - e_0 \cos (\phi - \omega_0 - \omega' \phi ) - \frac{1}{4} \hat{\alpha}_1 \Delta  \left (\frac{m}{a} \right )^{1/2} \frac{w}{\omega'}
\left ( \hat{w}_\perp \cos \phi - \hat{w}_\Omega \sin \phi \right )
\nonumber \\
& \quad
+ \frac{e_0}{4} \hat{\alpha}_2 w^2 \frac{\sin \omega' \phi}{\omega'} \left [ 2 \hat{w}_{\Omega} \hat{w}_{\perp} \cos( \phi + \omega_0) 
+ \left ( \hat{w}_{\perp}^2 - \hat{w}_{\Omega}^2 \right ) \sin(\phi + \omega_0) \right ] .
\label{eq8:rPFfinal}
\end{eqnarray} 
The first term in Eq.\ (\ref{eq8:rPFfinal}) is the normal contribution to $r/a$ resulting from the small eccentricity $e_0$, with the pericenter advancing at a rate $\omega'$.  The second term is a forced eccentricity of the orbit, with an amplitude proportional to $( \hat{w}_{\perp}^2 + \hat{w}_{\Omega}^2)^{1/2} = (1 - \hat{w}_h^2)^{1/2} \equiv \sin \psi$, where $\psi$ is the angle between the orbital angular momentum $\hat{\bm{h}}$ and the velocity $\bm{w}$ relative to the preferred frame, and a phase given by $\tan^{-1} (-\hat{w}_\Omega / \hat{w}_\perp )$ \cite{1992PhRvD..46.4128D}.   This effect is present even in the limit $e_0 \to 0$.  The final term is also a polarization of the orbit, proportional to $e_0$, with an amplitude modulated by the factor $\sin  \omega' \phi/\omega'$.   However, it vanishes in the limit $e_0 \to 0$.  

The other important effect of the preferred-frame perturbations is to cause the orbital angular momentum to precess.  Since $\hat{\bm h} = \sin \iota (\sin \Omega \bm{e}_x - \cos \Omega \bm{e}_y ) + \cos \iota \bm{e}_z$, variations in $\hat{\bm h}$ are given by
\begin{equation}
\Delta \hat{\bm h}  = \sin \iota \Delta \Omega \bm{e}_\Omega - \Delta \iota \bm{e}_\perp \,.
\end{equation}
Inserting the expressions for $\Delta \iota$ and $\Delta \Omega$ from Eqs.\ (\ref{eq8:deltaelements1}), taking the small $e$ limit, and noting that $\bm{e}_\Omega \cos \omega + \bm{e}_\perp \sin \omega = \bm{e}_{\rm P}$ and that $A_\perp \bm{e}_\Omega - A_\Omega \bm{e}_\perp = \bm{A} \times \hat{\bm h}$ we obtain
\begin{eqnarray}
\Delta \hat{\bm h}   &= - \frac{\pi}{2} \hat{\alpha}_1 \Delta \left (\frac{m}{a} \right )^{1/2} w_{\rm h} e \bm{e}_{\rm P}
-\pi \hat{\alpha}_2 w_h (\bm{w} \times  \hat{\bm h} )  \,,
\label{eq8:dhdt}
\end{eqnarray}
leading to a precession of the angular momentum vector $\bm{h}$.  

Searches for eccentricities induced by the $\hat{\alpha}_1$ term in Eq.\ (\ref{eq8:rPFfinal}) resulted in bounds on $ \hat{\alpha}_1$  as small as a few parts in $10^5$ \cite{2000ASPC..202..113W,2012CQGra..29u5018S}.   The tightest bound used a specific binary pulsar J1738+0333, whose orbit around its white-dwarf companion has an eccentricity $3.4 \times 10^{-7}$.  The analysis was helped by the fact that the white dwarf is bright enough to be observed spectroscopically, leading to accurate determinations of the key orbital parameters.  Furthermore, because the pericenter advances at a rate of about $1.6 \, {\rm deg \, yr}^{-1}$, the decade-long data span made it possible to partially separate any induced eccentricity, whose direction is fixed by the direction of $\bm{w}$, from the natural eccentricity, which rotates with the pericenter.  For this system the result was $|\hat{\alpha}_1| < 3.4 \times 10^{-5}$ \cite{2012CQGra..29u5018S}.

Limits on $\hat{\alpha}_2$  were obtained by looking for the precession of the orbital plane of a binary  system [see Eq.\ (\ref{eq8:dhdt})].   Such a precession would lead to a variation in the ``projected semimajor axis'' of the pulsar, $a_p \sin \iota$, a quantity that is measured very accurately in binary pulsar timing.   Combining data from the two wide-binary millisecond pulsar systems J1738+0333 and 
J1012+5307, Shao and Wex \cite{2012CQGra..29u5018S,2013CQGra..30p5020S} obtained the bound $|\hat{\alpha}_2 | < 1.8 \times 10^{-4}$.    

A given theory of gravity can be constrained or ruled out by combining these bounds on $\hat{\alpha}_1$ and $\hat{\alpha}_2$ with Eqs.\ (\ref{eq10:relativeacc4}), together with estimates of the sensitivities of any compact bodies in the system.  Those sensitivities will depend on both the theory and the equation of state of nuclear matter.

Returning to the full $N$-body Lagrangian (\ref{eq10:lagrangian1}), and working at quasi-Newtonian order, we can derive the leading signal of the failure of the universality of free fall (Nordtvedt effect) in a hierarchical three-body system, such as the pulsar J0337+1715 in a triple system with two white dwarf companions.   For a  two-body system in the
presence of a third body, the equations of motion become
\begin{eqnarray}
{\bm a}_1 &= -
{\cal G}_{12} m_2 \frac{\bx_{12}}{r_{12}^3} 
- {\cal G}_{13} m_3 \frac{\bx_{13}}{r_{13}^3} \,,
\nonumber \\
{\bm a}_2 &= {\cal G}_{12} m_1 \frac{\bx_{12}}{r_{12}^3} 
- {\cal G}_{23} m_3 \frac{\bx_{23}}{r_{23}^3}   \,.
\end{eqnarray}  
Following the method described for example in Sec.\ 13.3.3 of \cite{2014grav.book.....P}, it is straightforward to show that, for nearly circular coplanar orbits, the perturbation of the inner orbit induced by the Nordtvedt effect is given by 
\begin{equation}
\delta r = - \hat{\eta}_N \frac{R}{a_0} 
\frac{\omega_b^2(1+2\omega_b/\Lambda)}
{\omega_b^2 - \Lambda^2} \cos(\Lambda t + \Phi)  \,,
\label{eq14:deltarfinal}
\end{equation}
where $a_0$ and $R$ are the semimajor axes of the inner and outer orbits, respectively, $\omega_b$ is the angular frequency of the inner orbit, and $\Lambda \equiv \omega_b - \omega_3$ is the difference between the inner and outer orbit angular frequencies.   The ``strong-field'' Nordtvedt parameter $\hat{\eta}_N$ is given by
\begin{equation}
\hat{\eta}_N \equiv {\cal G}_{12} - {\cal G}_{13} \,.
\end{equation}
For the pulsar J0337+1715 we can ignore the sensitivities of the two white-dwarf companions, and write, for the specific cases of scalar-tensor and Einstein-{\AE}ther theories,
\begin{equation}
\hat{\eta}_N =  \Biggl \{ \begin{array}{cl} - \zeta s_1 &: {\rm scalar-tensor} \\
                           \hfill s_1/(1-s_1) &: {\rm Einstein-{\AE}ther} \,,
                         \end{array}  
\end{equation}
where we have set $G_N=1$ in Einstein-{\AE}ther theory.   Bounds on the Nordtvedt effect signal obtained from the data  can then be used to constrain these specific theories.

\section{Concluding remarks}
\label{sec:conclusions}

We have extended the modified EIH framework to incorporate the possibility of preferred-frame effects, giving a direct link between bounds on such effects derived from observations of binary pulsar systems and the fundamental parameters of alternative theories of gravity.  In contrast to the simplicity of the PPN formalism, the link between observation and theory here depends on the internal structure (sensitivities) of the compact bodies in the system.     

\ack
This work was supported in part by the National Science Foundation,
Grant No.\ PHY 16--00188.   We are grateful to Enrico Barausse, Guillaume Faye and Viraj Sanghai for useful discussions.

\appendix

\section{The compact-body Lagrangian formalism of Nordtvedt}

Nordtvedt's \cite{1985ApJ...297..390N} compact body formalism introduced two kinetic parameters $\delta M_a^{(2)}$, and $\delta M_a^{(4)}$, six symmetric parameters, $\Gamma_{ab}$, $\Theta_{ab}$, $\tau_{ab}$,  $\sigma_{ab}$, and $\xi_{ab}$, two antisymmetric parameters $\Phi_{ab}$ and $\Psi_{ab}$, and one three-body parameter, $\Gamma_{abc}$, symmetric on the second two indices.   Our mass rescaling argument implies that $\delta M_a^{(2)}=0$, and our rejection of Whitehead terms implies that 
$\xi_{ab} = \Psi_{ab} = 0$.   The remaining parameters are related to our modified EIH parameters by
\begin{eqnarray}
{\cal G}_{ab} & = (m_a m_b)^{-1} \Gamma_{ab} \,,
\nonumber \\
{\cal B}_{ab} &=\frac{1}{3}  (m_a m_b)^{-1} \left ( \Gamma_{ab} + 2\gamma \Theta_{ab} - 4 \Phi_{ab} \right ) \,,
\nonumber \\
{\cal D}_{abc} &= (m_a m_b m_c)^{-1} (2\beta-1)\Gamma_{abc} \,,
\nonumber \\
{\cal A}_a & = m_a^{-1} \delta M_a^{(4)} \,,
\nonumber \\
{\cal C}_{ab} &= (m_a m_b)^{-1} \left (\alpha_1 \tau_{ab} - 2\alpha_2 \sigma_{ab} \right )\,,
\nonumber \\
{\cal E}_{ab} &= 2 (m_a m_b)^{-1} \alpha_2 \sigma_{ab} \,,
\end{eqnarray}
where $\gamma$, $\beta$, $\alpha_1$ and $\alpha_2$ are the standard PPN parameters.

\section*{References}

\bibliographystyle{iopart-num}
\bibliography{refsTEGP2}

\providecommand{\newblock}{}
\begin{thebibliography}{10}
\expandafter\ifx\csname url\endcsname\relax
  \def\url#1{{\tt #1}}\fi
\expandafter\ifx\csname urlprefix\endcsname\relax\def\urlprefix{URL }\fi
\providecommand{\eprint}[2][]{\url{#2}}

\bibitem{LorentzDroste}
Lorentz H~A and Droste J 1917 {\em Versl.\ K.\ Akad.\ Wetensch.\ Amsterdam\/}
  {\bf 26} 392 english translation in Lorentz, H.A. 1937. Collected papers,
  Vol. 5, edited by Zeeman, P. and Fokker, A.D. Martinus Nijhoff

\bibitem{1916MNRAS..77..155D}
de~Sitter W 1916 {\em Mon. Not. R. Astron. Soc.\/} {\bf 77} 155--184

\bibitem{1965npga.book.....L}
{Levi-Civita} T 1965 {\em {The n-body problem in general relativity.}\/}
  (Dordrecht, Holland: D.~Reidel)

\bibitem{1964tstg.book.....F}
{Fock} V~A 1964 {\em {The theory of space, time and gravitation.}\/} (New York:
  Macmillan)

\bibitem{1938AnMat..39...65E}
{Einstein} A, {Infeld} L and {Hoffmann} B 1938 {\em Ann. Math.\/} {\bf 39}
  65--100

\bibitem{1968PhRv..169.1017N}
{Nordtvedt} Jr K 1968 {\em Phys. Rev.\/} {\bf 169} 1017--1025

\bibitem{1971ApJ...163..611W}
{Will} C~M 1971 {\em \apj\/} {\bf 163} 611

\bibitem{2007PhRvD..75l4025M}
Mitchell T and Will C~M 2007 {\em Phys. Rev. D\/} {\bf 75} 124025
  (\textit{Preprint} \eprint{0704.2243})

\bibitem{1975PhRvD..12.2183D}
{D'Eath} P~D 1975 {\em \prd\/} {\bf 12} 2183--2199

\bibitem{1978GReGr...9..809R}
{Rudolph} E and {B\"orner} G 1978 {\em Gen. Relativ. Gravit.\/} {\bf 9}
  809--820

\bibitem{1978GReGr...9..821R}
{Rudolph} E and {B\"orner} G 1978 {\em Gen. Relativ. Gravit.\/} {\bf 9}
  821--833

\bibitem{1980PhRvD..22.1853K}
{Kates} R~E 1980 {\em \prd\/} {\bf 22} 1853--1870

\bibitem{1985PhRvD..31.1815T}
{Thorne} K~S and {Hartle} J~B 1985 {\em \prd\/} {\bf 31} 1815--1837

\bibitem{1987PhRvD..36.2301A}
{Anderson} J~L 1987 {\em \prd\/} {\bf 36} 2301--2313

\bibitem{2000PhRvD..62f4002I}
{Itoh} Y, {Futamase} T and {Asada} H 2000 {\em \prd\/} {\bf 62} 064002
  (\textit{Preprint} \eprint{gr-qc/9910052})

\bibitem{2008PhRvD..78h4016T}
{Taylor} S and {Poisson} E 2008 {\em \prd\/} {\bf 78} 084016 (\textit{Preprint}
  \eprint{0806.3052})

\bibitem{tegp}
Will C~M 1981 {\em Theory and Experiment in Gravitational Physics\/}
  (Cambridge; New York: Cambridge University Press)

\bibitem{2014LRR....17....2B}
{Blanchet} L 2014 {\em \lrr\/} {\bf 17} 2 (\textit{Preprint}
  \eprint{1310.1528})

\bibitem{1975ApJ...196L..59E}
Eardley D~M 1975 {\em Astrophys. J. Lett.\/} {\bf 196} L59--L62

\bibitem{2010PhRvD..81h4060G}
{Gralla} S~E 2010 {\em \prd\/} {\bf 81} 084060 (\textit{Preprint}
  \eprint{1002.5045})

\bibitem{2013PhRvD..87j4020G}
{Gralla} S~E 2013 {\em \prd\/} {\bf 87} 104020 (\textit{Preprint}
  \eprint{1303.0269})

\bibitem{2001PhRvD..64b4028J}
{Jacobson} T and {Mattingly} D 2001 {\em \prd\/} {\bf 64} 024028
  (\textit{Preprint} \eprint{gr-qc/0007031})

\bibitem{2002cls..conf..331M}
{Mattingly} D and {Jacobson} T 2002 {Relativistic gravity with a dynamical
  preferred frame} {\em CPT and Lorentz Symmetry\/} ed {Kosteleck{\'y}} V~A
  (Singapore: World Scientific) pp 331--335 (\textit{Preprint}
  \eprint{gr-qc/0112012})

\bibitem{2010PhRvL.104r1302B}
{Blas} D, {Pujol{\`a}s} O and {Sibiryakov} S 2010 {\em Phys. Rev. Lett.\/} {\bf
  104} 181302 (\textit{Preprint} \eprint{0909.3525})

\bibitem{2011JHEP...04..018B}
{Blas} D, {Pujol{\`a}s} O and {Sibiryakov} S 2011 {\em J. High Energy Phys.\/}
  {\bf 4} 18 (\textit{Preprint} \eprint{1007.3503})

\bibitem{2004PhRvD..70h3509B}
{Bekenstein} J~D 2004 {\em \prd\/} {\bf 70} 083509 (\textit{Preprint}
  \eprint{astro-ph/0403694})

\bibitem{2008PhRvD..77l3502S}
{Skordis} C 2008 {\em \prd\/} {\bf 77} 123502 (\textit{Preprint}
  \eprint{0801.1985})

\bibitem{2006JCAP...03..004M}
{Moffat} J~W 2006 {\em \jcap\/} {\bf 3} 004 (\textit{Preprint}
  \eprint{gr-qc/0506021})

\bibitem{2000ASPC..202..113W}
{Wex} N 2000 {Small-eccentricity binary pulsars and relativistic gravity} {\em
  IAU Colloq. 177: Pulsar Astronomy - 2000 and Beyond\/} ({\em Astronomical
  Society of the Pacific Conference Series\/} vol 202) ed {Kramer} M, {Wex} N
  and {Wielebinski} R p 113 (\textit{Preprint} \eprint{gr-qc/0002032})

\bibitem{2012CQGra..29u5018S}
Shao L and Wex N 2012 {\em Class. Quantum Grav.\/} {\bf 29} 215018
  (\textit{Preprint} \eprint{1209.4503})

\bibitem{2013CQGra..30p5020S}
Shao L and Wex N 2013 {\em Class. Quantum Grav.\/} {\bf 30} 165020
  (\textit{Preprint} \eprint{1307.2637})

\bibitem{2014Natur.505..520R}
Ransom S~M, Stairs I~H, Archibald A~M, Hessels J~W~T, Kaplan D~L, van Kerkwijk
  M~H, Boyles J, Deller A~T, Chatterjee S, Schechtman-Rook A, Berndsen A, Lynch
  R~S, Lorimer D~R, Karako-Argaman C, Kaspi V~M, Kondratiev V~I, McLaughlin
  M~A, van Leeuwen J, Rosen R, Roberts M~S~E and Stovall K 2014 {\em Nature\/}
  {\bf 505} 520--524 (\textit{Preprint} \eprint{1401.0535})

\bibitem{tegp2}
{Will} C~M 2018 {\em Theory and Experiment in Gravitational Physics\/} 2nd ed
  (Cambridge, UK: Cambridge University Press) in press

\bibitem{2014LRR....17....4W}
{Will} C~M 2014 {\em \lrr\/} {\bf 17} 4 (\textit{Preprint} \eprint{1403.7377})

\bibitem{1985ApJ...297..390N}
{Nordtvedt} Jr K 1985 {\em \apj\/} {\bf 297} 390--404

\bibitem{1967RSPSA.298..123C}
{Chandrasekhar} S and {Contopoulos} G 1967 {\em Proc. R. Soc. A\/} {\bf 298}
  123--141

\bibitem{2013PhRvD..87h4070M}
Mirshekari S and Will C~M 2013 {\em Phys. Rev. D\/} {\bf 87} 084070
  (\textit{Preprint} \eprint{1301.4680})

\bibitem{2007PhRvD..76h4033F}
{Foster} B~Z 2007 {\em \prd\/} {\bf 76} 084033 (\textit{Preprint}
  \eprint{0706.0704})

\bibitem{2014PhRvD..89h4067Y}
{Yagi} K, {Blas} D, {Barausse} E and {Yunes} N 2014 {\em \prd\/} {\bf 89}
  084067 (\textit{Preprint} \eprint{1311.7144})

\bibitem{2014grav.book.....P}
{Poisson} E and {Will} C~M 2014 {\em {Gravity: Newtonian, Post-Newtonian,
  Relativistic}\/} (Cambridge, UK: Cambridge University Press)

\bibitem{1992PhRvD..46.4128D}
{Damour} T and {Esposito-Far\`ese} G 1992 {\em \prd\/} {\bf 46} 4128--4132

\end{thebibliography}

\end{document}